\documentclass[aps,prb,showpacs,twocolumn]{revtex4}
\usepackage{graphicx}
\usepackage{amsmath}
\usepackage{amssymb}
\usepackage{epsfig}
\usepackage{multirow}
\usepackage{dcolumn}
\usepackage{bm}

\begin{document}

\title{Competition between magnetic field dependent band structure and coherent backscattering
 in multiwall carbon nanotubes}
\author{B.~Stojetz, S.~Roche$^\dag$, C.~Miko$^\ddag$,
 F.~Triozon$^\dag$, L.~Forr\'{o}$^\ddag$, C.~Strunk\\
\it\small Institute of Experimental and Applied
Physics, University of Regensburg, D-93040 Regensburg, Germany\\
\it\small $^\dag$Commissariat \`a l'\'Energie Atomique, DRFMC/SPSMS, F-38054 Grenoble, France\\
 \it\small  $^\ddag$Institute of Physics of Complex Matter, FBS Swiss
Federal Institute of Technology (EPFL), CH-1015 Lausanne,
Switzerland}

\date{3 July 2006}

\begin{abstract}
Magnetotransport measurements in large diameter multiwall carbon
nanotubes (20-40 nm) demonstrate the competition of a magnetic-field
dependent bandstructure and Altshuler-Aronov-Spivak oscillations. By
means of an efficient capacitive coupling to a backgate electrode,
the magnetoconductance oscillations are explored as a function of
Fermi level shift. Changing the magnetic field orientation with
respect to the tube axis and by ensemble averaging, allows to
identify the contributions of different Aharonov-Bohm phases. The
results are in qualitative agreement with numerical calculations of
the band structure and  the conductance.

\end{abstract}
\pacs{73.63Fg, 72.80.Rj}
 \maketitle

The growing interest on molecular scale electronic transport has
motivated much research on carbon nanotubes\cite{DekkerReview}.
Single-wall carbon nanotubes with small diameters can be almost free
from contaminations or defects, making them ballistic 1D conductors.
In contrast, multiwall carbon nanotubes (MWNTs), with diameter
typically one order of magnitude larger, contain a much larger
density of defects and a reduced conductance. Thus, MWNTs represent
interesting model systems to investigate quantum interference
phenomena resulting from intrinsic disorder\cite{Schoenenberger},
and since a few years much effort has been devoted to understand
their quantum transport properties at low temperatures.

In particular, the cylindrical topology of the MWNTs allows to
explore unique conductance patterns in presence of external magnetic
fields. For weakly disordered systems and coaxial magnetic fields,
Altshuler-Aronov-Spivak (AAS) oscillations of the magnetoconductance
(MC) with a magnetic flux period of $h/2e$, are expected from the
theory of diffusive quantum transport\cite{Altshuler_Aronov_Spivak},
and have been indeed observed early on\cite{BachtoldAB}. Such
oscillations are explained by a field-modulated quantum interference
between clockwise and counterclockwise backscattered electronic
pathways, which encircle the cylinder. Since all such paths gain the
same Aharonov-Bohm phase per revolution, the interference is
modulated periodically with the magnetic flux. In perpendicular
magnetic field, periodic oscillations do not occur, but the same
coherent backscattering mechanism manifests itself in the phenomenon
of weak localization (WL).

Additionally, a remarkable magnetic field dependence of the
bandstructure of carbon nanotubes was predicted first by Ajiki and
Ando\cite{Ando}, resulting in a $h/e$-periodic band-gap oscillation
in a parallel magnetic field, associated to periodic splitting of
the van-Hove singularities in the density of states
(DoS)\cite{RDDS}. Recently, first indications of these bandstructure
effects have been seen experimentally by means of
photoluminescence\cite{Kono}, Coulomb
blockade\cite{jarillo,Bezryadin}, and Fabry-Perot type
interference\cite{Dai}. In Ref.~\onlinecite{Bezryadin} a
$h/e$-periodicity of single particle states has been reported, but
the subband spacing extracted was one order of magnitude too small.

However to date, the precise interpretation of the
magnetoconductance remains an issue of debate and controversies.
Indeed, while first reports of magnetotransport in MWNTs disregarded
any bandstructure contributions\cite{Schoenenberger}, others have
exclusively assigned Aharonov-Bohm oscillations to gap-modulations
and van-Hove singularities shifting\cite{WLControversy}. Fujiwara
and coworkers\cite{WLorientation} further show that a switching from
negative to positive magnetoresistance was possible by changing the
orientation of the magnetic field from parallel to perpendicular
direction with respect to the tube axis.

Unusual magnetotransport phenomena due to the coexistence of
different types of Aharonov-Bohm phases were then theoretically
addressed\cite{RS}. Recent experimental studies in small-diameter
SWNT and MWNT \cite{fedorov,13bis} have also reported on such
switching of the sign of the MC under gate voltage, but due to the
very small diameter, no Aharonov-Bohm oscillations could be
resolved.

In this Letter, the interplay between the different magnetic flux
effects and the bandstructure and localization phenomenona is
investigated by studying the MC in thick MWNTs over a wide range of
Fermi level positions. In contrast to earlier works, the large
diameter (29 - 35nm) allows a threading flux $\phi\sim 3\phi_0$ at
magnetic fields of 16 Tesla ($\phi_0=h/e$ the quantum flux). In
addition, the transport through several several subbands is explored
by virtue of a strongly coupled Al backgate electrode. The measured
MC reveals clear signatures of coexisting magnetic field-dependent
band structure and coherent backscattering. The $h/e$-periodic
contribution agrees well with numerical calculations of the DoS,
while the $h/2e$-contribution is clearly of AAS-type. In contrast, a
perpendicular field only weakly affects the DoS, the
magnetoresistance being fully dominated by the coherent
backscattering.

\begin{figure}
\includegraphics[width=85mm]{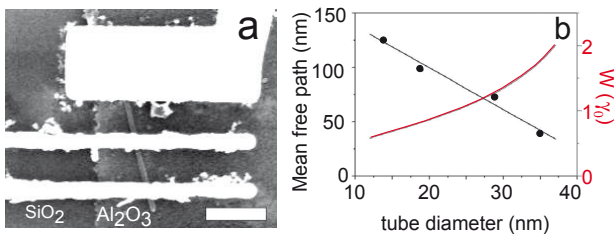}
\caption{\label{fig0}a) Scanning electron micrograph of sample
  D. An individual MWNT is deposited on top of an oxidized Al film and
  contacted by three Au fingers. Only the upper two fingers are
  used. The scalebar corresponds to 500 nm. b) Estimate of the elastic mean free path
  $\ell_{\rm el}$ at the CNP (black dots) and effective Anderson disorder $W$ (red)
  vs. tube diameter at 10 K (see text). The black line represents a linear fit to $\ell_{\rm el}$.}
\end{figure}

The samples were produced by deposition of individual MWNTs on
oxidized Al strips and subsequent fabrication of Au contacts
(\cite{Stojetz,Stojetz2} for details). The two-terminal conductance
was measured with conventional lock-in techniques, while a dc
voltage, $U_{\rm gate}$, was applied to the Al backgate. Four
samples, A to D, were used with diameters $d_{\rm tube}$ of 14, 19,
29 and 35 nm, respectively. A scanning electron micrograph of sample
D is shown in Fig.1a.

The basic parameter determining the electric transport properties
besides the band structure is the elastic mean free path $\ell_{\rm
el}$. In carbon nanotubes, $\ell_{\rm el}$ has been found to depend
strongly on the position of the Fermi energy $E_F$ \cite
{Stojetz,Triozonlel}, with a pronounced maximum in the vicinity of
the charge neutrality point (CNP), where bonding and anti-bonding
bands cross. In a conductor of length $L$ with $N_{\small\perp}$
conducting channels per spin direction, the two terminal resistance
$R$ is given by
\begin{equation}
R=R_C+\frac{h}{2e^2}\left( \frac{N_{\small\perp}\ell_{\rm
el}}{L+\ell_{\rm
      el}}-\frac{L_\varphi}{L}\right)^{-1},\label{eq:lel}
\end{equation}
where $L_\varphi$ is the phase coherence length and $R_C$ is the
contact resistance 
and $N_{\small\perp}=2$, at the CNP of a metallic nanotube. The
position of the CNP can be identified by means of a strongly coupled
Al gate and is indicated by a shallow maximum in $R(V_{gate})$ at
\mbox{300 K}. As in\cite{Stojetz}, the corresponding $R$(CNP) can be
inferred. By assuming a typical value for $R_C$ of 2.5 k$\Omega$, a
lower bound to $\ell_{\rm el}$ at the CNP can be estimated from
Eq.~\ref{eq:lel}. The result as a function of tube diameter is
presented in Fig.1b. This lower bound for $\ell_{\rm el}$ decreases
nearly linearly with $d_{\rm tube}$, from 130 nm to 40 nm for
diameters of 14 nm and 35 nm, respectively. As soon as higher
nanotube subbands are occupied, Eq.\ref{eq:lel} as well as numerical
results\cite{Triozonlel} predict a strong decrease of $\ell_{\rm
el}$ by about one order of magnitude. With the exception of a narrow
region around the CNP, all of our MWNTs are thus in the diffusive
regime. Only at the CNP $\ell_{\rm el}$ can exceed both $d_{\rm
tube}$ and $L_\varphi$.

In order to estimate an effective strength of the disorder
potential, we use the analytical form of $\ell_{\rm el}$ that has
been derived for Anderson disorder with width $W$, i.e.,
\mbox{$(W/\gamma_0)^2=(18\pi/\sqrt{3})\times (d_{\rm tube}/\ell_{\rm
el})$}, where $\gamma_0\simeq$3~eV is the C-C hopping matrix
element\cite{Triozonlel}. The result is presented in Fig.\ref{fig0}b
(red curve). $W$ increases with diameter from below 1~eV to 2~eV.
Thus, the disorder apparently increases with diameter for MWNTs,
probably due to a larger defect density in larger diameter MWNTs, as
often reported experimentally\cite{Millie}. In the following, in
agreement with prior studies
\cite{Schoenenberger,BachtoldAB,Stojetz}, the transport is assumed
to be restricted to the outermost conducting shell\cite{Refnew}.

\begin{figure}
\includegraphics[width=85mm]{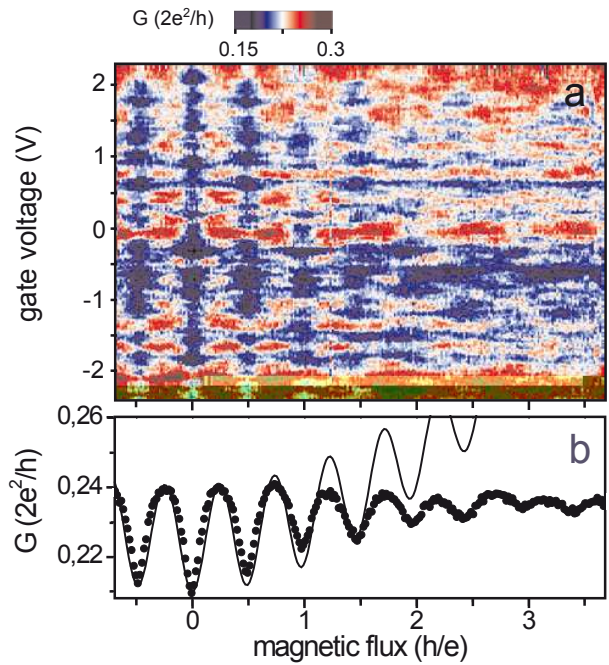}
\caption{\label{fig1}a) Color representation of the
  differential conductance as a function of gate voltage $U_{\rm
  gate}$ and coaxial magnetic field at $4.5$ K. The CNP is located
  in the center of the horizontal conductance valley near \mbox{$U_{gate}=-0.6$~V}.
  b) Magnetoconductance averaged over all gate voltages (dots).
  The line is the best fit according to the AAS-theory. }
\end{figure}

Next, the conductance $G$ of samples C and D was recorded as a
function of gate voltage $U_{\rm gate}$ and \textit{coaxial}
magnetic field $B$ at a temperature of 4.5~K. The data for sample D
are presented in Fig.\ref{fig1}a in a color representation. The
conductance is strongly modulated as a function of both magnetic
field and gate voltage $U_G$. At our relatively high temperatures,
$T\gtrsim4.5\;$K, the behavior of $G$ is dominated by bandstructure
effects and universal conductance fluctuations. For each value of
the magnetic field, $G$ is minimal in the vicinity of $U_{\rm
gate}=-0.6$~V, which can be identified with the CNP from room
temperature measurements (not shown). Most MC traces show a minimum
at zero magnetic field as well as at multiples of $\Delta B=2.2$~T.
$\Delta B$ agrees well with a flux period of $h/2e$, which
corresponds to 2.2~T for $d_{\rm tube}=35$~nm. In addition, several
conductance features are repeated periodically with a field period
$2\Delta B$, e.g., the ridges of high conductance around zero gate
voltage. Thus, two periodicities $\Delta B$ and 2$\Delta B$ are
superimposed, which signals the coexistence of the $h/2e$
AAS-oscillations originating from coherent backscattering and $h/e$
Aharonov-Bohm oscillations of the bandstructure.

If the gate voltage range is sufficiently large, an average over all
MC traces removes the universal conductance fluctuations as well as
the bandstructure modulations \cite{Stojetz}. Only the contribution
of the coherent backscattering to the conductance is predicted to
remain\cite{Altshuler_Aronov_Spivak}. The ensemble average for
$U_{\rm G}$ in the interval [$-2.3$~V, $2.3$~V] is presented in
Fig.\ref{fig1}b. The average conductance $\langle G\rangle$
oscillates with a flux period of approximately $\simeq h/2e$
(corresponding to the outer diameter of the tube), as expected. The
solid line in Fig.\ref{fig1}b corresponds to the best fit of the AAS
theory \cite{Altshuler_Aronov_Spivak}. The free parameters were
$L_{\varphi}$ and the angle $\theta$ of the magnetic field with
respect to the tube axis. The best agreement is obtained for
$L_{\varphi}=30\;$nm and $\theta=2^\circ$. The quality of the fit is
satisfactory for small fields, while for larger fields the
oscillations are strongly damped and deviations in the parabolic
background occur. The origin of these deviations is not clear. The
flux periodicity of approximately $h/2e$ suggests that the current
is carried only by few outermost shells of the MWNT. Since the
periodicity reflects the average shell diameter of the transport,
the inner shells apparently do not  much contribute to the current.

\begin{figure}
\includegraphics[width=85mm]{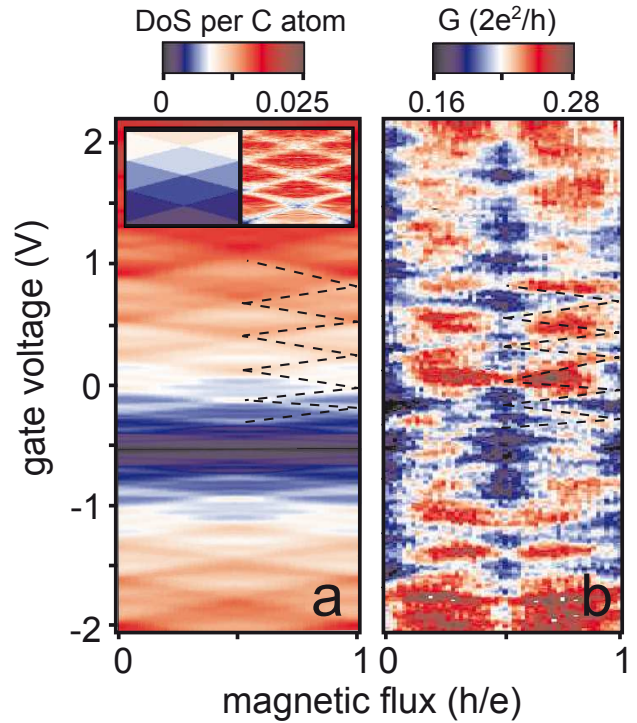}
\caption{\label{fig2}a) Color representation of the
  calculated DoS for a (260, 260)-singlewall
  nanotube as a function of gate voltage and coaxial
  magnetic flux. The dashed lines indicate ridges of high DoS induced by
  the $h/e$-periodic splitting of van Hove singularities. Insets:
  Conductance of a clean (left) and disordered, $W=\gamma_0/5$, (right)
(22, 22) nanotube (the x-scale is given in units of $h/e$, the
y-scale corresponds to a small number of subbands above CNP).  b)
First $h/e$-period of the data in
  Fig. \ref{fig1}a. Dashed lines indicate diamond like ridges of high conductance.}
\end{figure}

To unravel the precise contribution of bandstructure effects from
quantum interferences phenomena, a theoretical analysis of the DoS
and quantum conductance in clean and disordered nanotubes is
performed. First, in Fig.3a, the density of states of clean (260,
260) armchair nanotube (with diameter of 35nm) is shown as a function
of the gate voltage, $U_{\rm gate}$, and coaxial magnetic flux in the
interval [0, $h/e$]. For the comparison of the calculated DoS to the
measurement, the energy scale $E$ has to be converted into an
equivalent gate voltage scale. This has been done successfully
earlier by assuming a constant capacitive coupling $C_{\rm
gate}\simeq 300 \text{~aF}/\mu$m between the gate and the tube
\cite{Stojetz}. In a color-plot, a diamond-shaped structure is
clearly visible and manifest the splitting and shift of the van-Hove
singularity, driven by the magnetic field. The DoS increases from the
CNP to the higher electronic subbands. Some diamonds are highlighted
with dashed lines.

Figure~3b shows a close-up of the experimental data in Fig.~2a in
the magnetic flux range between 0 and $h/e$ (sample D). We find
conductance ridges, which are most pronounced at intermediate flux
values $\Phi=h/4e$ and $\Phi=3h/4e$, rather than at  $\Phi=0$ and
$\Phi=h/2e$ as expected from the AA-theory. The reason is the
suppression of the conductance at the latter values by the
AAS-effect. The latter is caused by the coherent backscattering in
the disorder potential, as illustrated by the ensemble average in
Fig.~2b. Taking his into account, we note are many gate voltages
(e.g. 1.8, 1.3, 0.7, 0.4, 0.2, -0.3, -0.9, -1.8~V), where a
conductance valley at  $\Phi=h/2e$ is surrounded by conductance
ridges (highlighted by dashed lines) in agreement with the
AA-prediction. Although these similarities are not sufficient for a
one-to-one assignment of calculation and experiment, they
nevertheless demonstrate that the position, the shape, and also the
size of the diamonds agree reasonably well with our simple model.
Similar results are obtained for sample C. In our previous
experiment\cite{Stojetz}, we could link peaks in the conductance
with the positions of van-Hove singularities in the DoS at the onset
of the one-dimensional subbands. A very similar link between the
peaks in the conductance and the DoS is reproduced by the
theoretical curves in Fig.~2a.

Next, the conductance of a clean and a disordered nanotube are
computed based on the Landauer-B\"uttiker framework \cite{TLR}.  Due
to computational limitations, we here restrict to a (22, 22)
nanotube (with 3~nm diameter), but conclusions applied equally to
larger diameter owing to scaling laws. In the clean system, the
ballistic conductance (Fig.~3a left inset) displays a diamond
structure identical to that of the DoS (not shown here), but gives
the modulation of available quantum channels under magnetic field
and Fermi level shift. The conductance is constant within
rhomb-shaped regions in the $B$-$E_F$-plane. At higher energy,
additional conductance channels are populated, which causes a
stepwise increase of the conductance.

To explore the effect of the Anderson disorder, one can extrapolate
from Fig.~1b an effective disorder strength that would correspond to
an experimental measurement for a (22, 22) nanotube. The obtained
value lies within $[1/10, 1/2]\gamma_{0}$. Using $W=\gamma_{0}/5$,
the average quantum conductance for disordered (22, 22) nanotubes are
computed (Fig.~3a right inset). The diamond shape pattern remains
robust for lower energies, while it progressively degrades as higher
energy subbands are involved in conduction. Therefore, the
conductance increase due to the opening of channels is
(over)compensated by enhanced backscattering. Calculations for much
larger disorder (in the order of twice the spectral bandwith) show
that such a degradation is enhanced, but rhomb-shaped regions still
survive close to CNP. The estimated localization lengths are slightly
smaller or of the order of the coherence length, confirming that we
are in the WL regime.

\begin{figure}
\includegraphics[width=85mm]{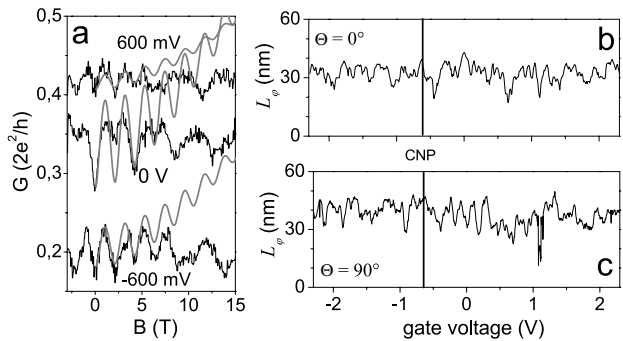}
\caption{\label{fig3}a) Magnetoconductance (MC) traces at $U_{\rm
gate}=$
  600 mV, 0V, -600 mV (top to bottom) (black) and fits of AAS theory
  (grey). Right panel: Phase coherence length vs. gate voltage for coaxial field
  (b) and perpendicular field (c). Vertical lines indicate the position of the CNP.}
\end{figure}

After rotating the sample by 90$^\circ$ at low temperatures the
conductance $G$ has been recorded as a function of gate voltage and
a magnetic field \textit{perpendicular} to the tube axis. For gate
voltages in the same range as for the prior case, the MC traces
exhibit a minimum at $B=0$ (not shown here), which is the negative
magnetoresistance effect of the WL. \cite{Altshuler_Aronov_Spivak}.
As expected for perpendicular magnetic fields, no field-periodic
oscillations of $G$ are observed. Instead, $G$ is maximal at
B$\approx$2~T. For higher fields, $G$ decreases at most gate
voltages. This could be explained in terms of bandstructure effects
in perpendicular fields, where a decrease of the DoS is predicted
close to the CNP\cite{RDDS}. Finally, in contrast to the
$h/e$-splitting of the van-Hove singularities in coaxial field, no
such effect is observed for perpendicular fields.

The relevant length scale governing the contribution of coherent
backscattering to the conductance is the phase coherence length
$L_{\varphi}$, which can be independently extracted from the fits of
$G(B,U_\text{gate})$ to the WL and AAS theories, respectively. The
result should be independent of the orientation of $B$ with respect
to the tube axis. Figure 4a shows some representative $G(B_{\|})$
traces at different $U_\text{gate}$. The fits reproduce the
oscillatory part of the the experimental traces, but contain also a
pronounced monotonic background, which is absent in the data. At
present, the origin of this discrepancy is unclear, since this
background is expected from the the small misalignment of 2$^\circ$
between the tube axis and the magnetic field. Nevertheless,
$L_\varphi$ can be extracted with sufficient accuracy from the
amplitude of the first period of the oscillation. The result is
plotted in Fig. 4b and shows pronounced variations with
$U_\text{gate}$. Figure 4c shows $L_\varphi(U_\text{gate})$
extracted from the $G(B_\bot)$-traces. The two
$L_{\varphi}(U_\text{gate})$ curves agree well in absolute value.
The correlation in the position of the minima appears to be weak.
Given the superimposed effect of aperiodic conductance fluctutations
(UCF) in the individual MC-traces and the systematic shift of the
van-Hove singularities with magnetic field for parallel orientation
this is not too surprising. Unlike in the case of samples A and B
\cite{Stojetz}, the disorder in the 35~nm thick sample D seems to be
too strong compared with the subband spacing, to allow an analogous
identification of all subband adjacent to the CNP.

In conclusion, unambiguous signatures of magnetic-field dependent and
quantum interference phenomena have been shown to jointly contribute
to the magneto conductance of large diameter MWNTs. Different types
of Aharonov-Bohm phases were identified, resulting in a comprehensive
description of quantum transport in diffusive multiwall carbon
nanotubes.


 We thank G.~Cuniberti, M.~Grifoni, N.~Nemec, K.~Richter,
and C.~Sch\"onen\-berger for inspiring discussions and C.~Mitzkus and
D.~Weiss for experimental support. Funding by the Deutsche
 Forschungs\-gemeinschaft within the Graduiertenkolleg 638 is
 acknowledged. The work in Lausanne was supported by the Swiss
 National Science Foundation.


\end{document}